\documentclass[aps,prb,twocolumn,superscriptaddress,showpacs]{revtex4-1}

\usepackage{graphicx}
\usepackage{calc}
\usepackage{bm}
\usepackage{color}
\usepackage{textcomp}

\begin{document}

\title{Reentrant proximity-induced superconductivity for GeTe semimetal}

\author{V.D.~Esin}
\author{D.Yu.~Kazmin}
\author{Yu.S.~Barash}
\author{A.V.~Timonina}
\author{N.N.~Kolesnikov}
\author{E.V.~Deviatov}

\affiliation{Institute of Solid State Physics of the Russian Academy of Sciences, Chernogolovka, Moscow District, 2 Academician Ossipyan str., 142432 Russia}

\date{\today}

\begin{abstract}
  We experimentally investigate charge transport in In-GeTe and In-GeTe-In proximity devices, which are formed as junctions between superconducting indium leads and thick single crystal flakes of $\alpha$-GeTe topological semimetal. We observe nonmonotonic effects of the applied external magnetic field, including reentrant superconductivity in In-GeTe-In Josephson junctions: supercurrent reappears at some finite magnetic field. For a single In-GeTe Andreev junction, the superconducting gap is partially suppressed in zero magnetic 
field, while the gap is increased nearly to the bulk value for some finite field before its full 
suppression. We discuss possible reasons for the results obtained, taking into account spin 
polarization of Fermi arc surface states in topological semimetal $\alpha$-GeTe with a strong 
spin-orbit coupling. In particular, the zero-field surface state spin polarization partially 
suppresses the superconductivity, while it is recovered due to the modified spin-split surface state configuration in finite fields. As an alternative possible scenario, the transition into the Fulde-Ferrell-Larkin-Ovchinnikov state is discussed. However, the role of strong spin-orbit coupling in forming the nonmonotonic behavior has not been analyzed for heterostructures in the Fulde-Ferrell-Larkin-Ovchinnikov state, which is crucial for junctions involving GeTe topological semimetal.
\end{abstract}

\pacs{73.40.Qv  71.30.+h}

\maketitle

\section{Introduction}

Recent interest to topology in condensed matter is mostly moved from topological insulators to 
topological semimetals~\cite{armitage}.  Crystal symmetry protects band crossing at some distinct Dirac points, so even the bulk spectrum is gapless in semimetals.  In Weyl semimetals (WSM) every crossing point splits  into two Weyl nodes with opposite chiralities due to the time reversal or inversion symmetries breaking. The projections of two Weyl nodes on the surface Brillouin zone are connected by a Fermi arc, which represents the topologically protected surface state~\cite{armitage}. Fermi arcs have been experimentally demonstrated by angle-resolved photoemission spectroscopy (ARPES), e.g. for MoTe$_2$ and WTe$_2$ three-dimensional crystals~\cite{wang,wu}. Due to the spin-momentum locking, Weyl semimetals are considered as attractive materials for spin investigations~\cite{spin-burkov}. Spin- resolved ARPES measurements demonstrated nearly full spin polarization of Fermi arcs, e.g. in TaAs~\cite{lv2015,xu16} and in WTe$_2$ Weyl semimetals~\cite{li2017,soluyanov}.

While  Weyl semimetals with broken time-reversal symmetry are magnetically ordered 
materials~\cite{armitage,mag2,mag3,kagome}, WSMs with broken inversion symmetry 
have to obtain bulk ferroelectric polarization~\cite{armitage,TSreview}. Due to  the gapless bulk 
spectrum~\cite{armitage}, a non-centrosymmetric Weyl semimetal is the natural representation of 
the novel concept of  intrinsic polar metal~\cite{PM,pm1,pm2,pm4}. Among these materials, GeTe is of special interest~\cite{GeTespin-to-charge,GeTereview} due to the reported  giant Rashba band splitting~\cite{GeTerashba,Morgenstern,GeTerashba1}. The ferroelectric control of spin-to-charge conversion was shown in epitaxial GeTe films~\cite{GeTespin-to-charge}, also, direct correlation between ferroelectricity and spin textures was demonstrated in this material~\cite{spin text}.

Ferroelectric Rashba semiconductor GeTe is predicted to be topological semimetal in low-temperature (below 700~K) ferroelectric $\alpha$-phase~\cite{ortix}.  The bulk Dirac points evolve either into pairs of Weyl nodes or into mirror-symmetry protected nodal loops upon breaking inversion symmetry~\cite{ortix}. Moreover, ferroelectric $\alpha$-GeTe is unveiled 
to exhibit an intriguing multiple nontrivial topology of the electronic band structure due to the 
existence of triple-point and type-II Weyl fermions~\cite{triple-point}. Band structures with triple band crossing are rather rare in condensed matter. This results, e.g., in a complicated spin texture around the triple point~\cite{triple-point}.

The direct measurement of the Rashba-split surface states of $\alpha$-GeTe(111) has been 
experimentally realized thanks to K doping~\cite{GeTesurfStates}. It has been shown that the 
surface states are not the result of band bending and that they are decoupled from the bulk states. The giant Rashba splitting of the surface states of $\alpha$-GeTe is largely arising from the inversion symmetry breaking in the bulk~\cite{GeTesurfStates}.

Topological materials  exhibit non-trivial physics in  proximity with a superconductor. For the
topological insulators, it is expected to  allow topological 
superconductivity  regime~\cite{Fu}, which stimulates a search for  Majorana 
fermions~\cite{reviews}. Zero-energy or near zero-energy (non-topological) surface 
Andreev states can appear in such heterostructures, see, for 
example~\cite{Lutchynetal,DengetalMarcus}. Also, field-induced reentrant effects have  been identified in uniform materials  with combined magnetization and strong spin-orbit coupling,  such as the uranium-based superconductors~\cite{Levyetal2005,Miyakeetal2008,Mineev2017,Machida2021,Kinjoetal2023,Rosueletal2023}. 
The reentrant behavior is usually associated here with the influence of the 
external magnetic field  on the spin subsystem, possibly, close to the tricritical point or a metamagnetic transition. Thus, it it reasonable to study also proximity-induced superconductivity in $\alpha$-GeTe. 

Here, we experimentally investigate charge transport in In-GeTe and In-GeTe-In proximity devices, which are formed as junctions between superconducting indium leads and thick single crystal flakes of $\alpha$-GeTe topological semimetal. We observe nonmonotonic effects of the applied external magnetic field, including reentrant superconductivity in In-GeTe-In Josephson junctions: supercurrent reappears at some finite magnetic field. For a single In-GeTe Andreev junction, the superconducting gap is partially suppressed in zero magnetic field, while the gap is increased nearly to the bulk value for some finite field before its full suppression. We discuss possible reasons for the results obtained, taking into account spin polarization of Fermi arc surface states in topological semimetal $\alpha$-GeTe with a strong spin-orbit coupling.

\section{Samples and technique}

\begin{figure}
\includegraphics[width=\columnwidth]{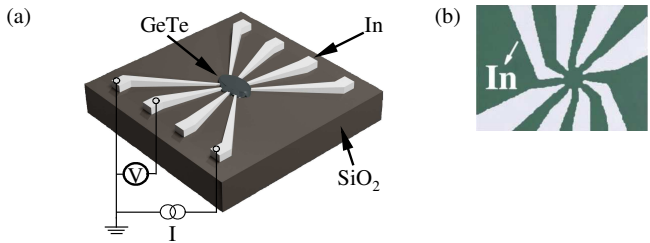}
\caption{(Color online) (a) A top-view image of the sample with In leads. A thick (1~$\mu$m)  mechanically exfoliated GeTe flake is placed on the  pre-defined In leads pattern to form multiple In-GeTe junctions. Electron transport is measured in a standard  three-point technique. An additional (fourth) wire to the grounded contact allows to exclude all the wire resistances. The advantage of the proposed three-point technique is that it it suitable for both the Andreev reflection and Josephson current investigations. (b) Image of the In leads pattern without a flake, to demonstrate the experimental geometry.  The leads are formed by lift-off technique after thermal evaporation of 100~nm In on the standard Si/SiO$_2$ substrate. 2~$\mu$m wide In leads are separated by 5~$\mu$m intervals.}
\label{fig1}
\end{figure}

GeTe single crystals were grown by physical vapor transport in the evacuated silica ampule. The initial GeTe load was synthesized by direct reaction of the high-purity (99.9999\%) elements in vacuum. For the  crystals growth, the initial GeTe load serves as a source of vapors: it was melted and kept at 770-780$^\circ$ C for 24 h. Afterward, the source was cooled down to 350$^\circ$ C at the 7.5 deg/h rate. The GeTe crystals grew during this process on the cold  ampule walls somewhat above the source. The GeTe composition is verified by energy-dispersive X-ray spectroscopy. The powder X-ray diffraction analysis confirms single-phase GeTe, the known structure model~\cite{GeTerashba} is also refined with single crystal X-ray diffraction measurements. The quality of our GeTe material was also tested in capacitance measurements~\cite{GeTecap}, the results confirmed giant Rashba splitting~\cite{GeTerashba} in our GeTe single crystals. 

Fig.~\ref{fig1}(a) shows a top-view image of a sample.  The topological semimetals are essentially three-dimensional objects~\cite{armitage}, so we have to select  relatively thick (above 1~$\mu$m) flakes. In this case,  the desired experimental geometry can not be defined by usual mesa etching, so the geometry is formed by  pre-defined contact pattern on the oxidized silicon substrate. GeTe single crystal flakes are obtained  by mechanical exfoliation from the initial ingot.   We choose the $\approx$100~$\mu$m wide flakes with defect-free surface by optical microscope. The chosen flake is placed immediately on the standard Si/SiO$_2$ substrate  with pre-defined 2~$\mu$m wide In contact leads. 

The leads pattern is prepared by lift-off technique after thermal evaporation of 100~nm In. The 100~nm thick, 2~$\mu$m wide In leads are separated by 5~$\mu$m intervals, as depicted in Fig.~\ref{fig1} (b). After initial single-shot pressing by another oxidized silicon substrate,  the flake is firmly connected to the In leads. This procedure provides transparent In-GeTe junctions (about 2~Ohm resistance), stable in different cooling cycles, which has been verified  before for a wide range of materials~\cite{cdas,inwte1,inwte2,incosns,aunite,timnal,infgt}. As an additional advantage, the obtained In-GeTe junctions are protected from any contamination by SiO$_2$ substrate. 

We study electron transport across a single In-GeTe junction in a standard three-point technique: one In contact is grounded, the neighbor (5~$\mu$m separated) In contact is used as a voltage probe, while current is fed through another (outlying) contact, as schematically presented in Fig.~\ref{fig1} (a). We use an additional (fourth) wire to the grounded contact,  so all the wire resistances are excluded, which is necessary for low-impedance samples.

The advantage of the proposed three-point technique is that it it suitable for both the Andreev reflection and Josephson current investigations. If there is no supercurrent between two neighbor In leads,  the potential probe mostly reflects the voltage drop across the grounded In-GeTe interface, i.e. Andreev reflection. In contrast, the potential probe shows exactly zero  if the two neighbor In probes are connected by the  Josephson current~\cite{inwte1,inwte2,incosns,aunite,infgt}.

To obtain $dV/dI(V)$ and $dV/dI(I)$ characteristics, dc current is additionally modulated by a low (100~nA) ac component. We measure both dc ($V$) and ac ($\sim dV/dI$) voltage components  with a dc voltmeter and a lock-in amplifier, respectively. The signal is confirmed to be independent of the modulation frequency within 100 Hz -- 10kHz range, which is defined by the applied filters.  The measurements below are performed within the 30~mK -- 1.2~K temperature range in a dilution refrigerator equipped with a superconducting solenoid.

\section{Experimental results}

\subsection{Single In-GeTe junctions}

\begin{figure}
\includegraphics[width=\columnwidth]{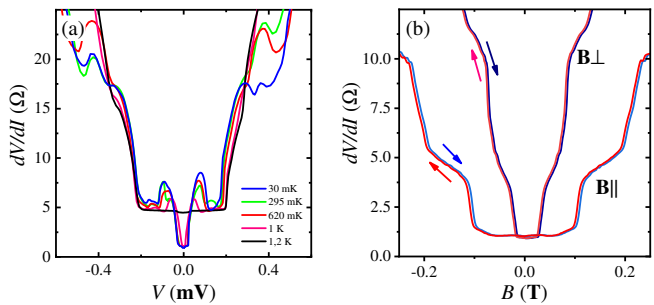}
\caption{(Color online) Andreev behavior of $dV/dI$ differential resistance for a single In-GeTe junction. (a) $dV/dI(V)$ curves in zero magnetic field at different temperatures. Differential resistance is diminished within $\pm 0.25$~mV bias interval, which is refined from the  magnetic field dependence (see the text). The obtained value is quite reasonable for the indium film on a top of a GeTe crystal with giant Rashba splitting~\cite{GeTerashba}. Temperature has low effect on  $dV/dI(V)$ curves, only the zero-bias anomaly appears below 1~K. (b) Magnetic field suppression of the Andreev reflection for normal and in-plane field orientations. There is no hysteresis for two opposite sweep directions. For the planar experimental geometry, it is natural to have the critical field anisotropy. The curves show two-step suppression, which appears by two kinks at $\approx5$~$\Omega$ and $\approx10$~$\Omega$ resistance levels. The first kink is the suppression of the zero-bias anomaly, while the second one belongs to the suppression of Andreev reflection. The $dV/dI(B)$ curves are obtained  at zero bias voltage and at the lowest 30~mK temperature. }
\label{fig2}
\end{figure}

As prepared, two neighbor In probes are not connected by supercurrent through $5 \mu$m distance along the GeTe surface, so the measured voltage drop reflects the resistance of a single (grounded)  In-GeTe junction in Fig.~\ref{fig1} (a). We verify that for a fixed grounded In contact,  the $dV/dI(V)$ curve is independent of the mutual positions of current/voltage probes, so it indeed  reflects the resistance of In-Gete interface  without noticeable admixture of the sample’s bulk. Temperature has low effect on  $dV/dI(I)$, only the zero-bias anomaly appears below 1~K, see Fig.~\ref{fig2} (a).

$dV/dI(V)$ differential resistance curves show well-developed Andreev behavior in Fig.~\ref{fig2} (a) and (b). Since Andreev reflection allows subgap transport of Cooper pairs,  it appears experimentally as the resistance drop for voltages within the superconducting gap~\cite{andreev,tinkham}.  As it can be seen in Fig.~\ref{fig2} (a), differential resistance is diminished within some bias interval, which is approximately twice smaller than the  known $0.5$~mV  bulk indium gap~\cite{indium}. The exact gap value can be refined from the  magnetic field dependence.

\begin{figure}
\includegraphics[width=\columnwidth]{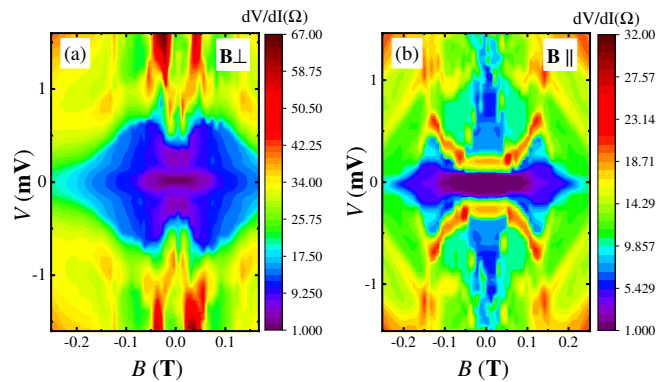}
\caption{(Color online) The detailed colormaps of the gap suppression for two, in-plane (a) and normal (b), magnetic field orientations, respectively. The suppression pattern is of butterfly shape, demonstrating non-monotonic behavior of the proximized superconductivity on the surface of GeTe single crystal. The superconducting gap is partially suppressed in zero magnetic field, while it is increased nearly to the bulk value for some finite field (0.05~T for normal field and 0.15~T for the in-plane one) before its full suppression. The data are obtained at 30~mK temperature. 
 }
\label{fig3}
\end{figure}

Andreev reflection can be suppressed by magnetic field~\cite{andreev,tinkham}, see Fig.~\ref{fig2} (b) for normal and in-plane field orientations, respectively. The curves are obtained at zero bias voltage and at the lowest 30~mK temperature. One can observe two-step suppression, as appears by two kinks at $\approx5$~$\Omega$ and $\approx10$~$\Omega$ resistance levels. The first kink is the suppression of the zero-bias anomaly, while the second one belongs to the suppression of Andreev reflection. The coefficient 2 between the levels confirms transparent In-GeTe interfaces in our experiment~\cite{tinkham}: the resistance drops in two times within the superconducting gap. Thus, we should accept  $\approx10$~$\Omega$ as the normal resistance value, which allows to clarify the superconducting gap value as 0.25~meV in Fig.~\ref{fig2} (a).  The obtained value is quite reasonable for the indium film on a top of a 
GeTe crystal with giant Rashba splitting~\cite{GeTerashba}, which is characterized by strong spin polarization of the surface~\cite{spin text,GeTesurfStates}. 

For the in-plane field, 
the zero-bias anomaly is suppressed at about 100~mT, while it is  smaller for the normal field orientation. The superconducting gap survives to higher fields, which are 0.2~T and 0.075~T respectively.  It is natural to have the critical field anisotropy in Fig.~\ref{fig2} (b) for  planar experimental geometry of our experiment.

The detailed picture of the gap suppression is quite complicated, see the colormaps in Fig.~\ref{fig3} (a) and (b) for two magnetic field orientations. While the zero-bias anomaly is suppressed monotonically (black color, about 1~$\Omega$ level), the gap (the blue regions) is sharply increased above some  magnetic field value, which is about 0.05-T for normal field and 0.15~T for the in-plane one, see the $\approx$10~$\Omega$ level in Fig.~\ref{fig3} (a) and (b). Thus, the suppression pattern is of butterfly shape, demonstrating non-monotonic behavior of the proximized superconductivity on the surface of GeTe single crystal.

\subsection{Double In-GeTe-In junctions} \label{double}

\begin{figure}
\includegraphics[width=\columnwidth]{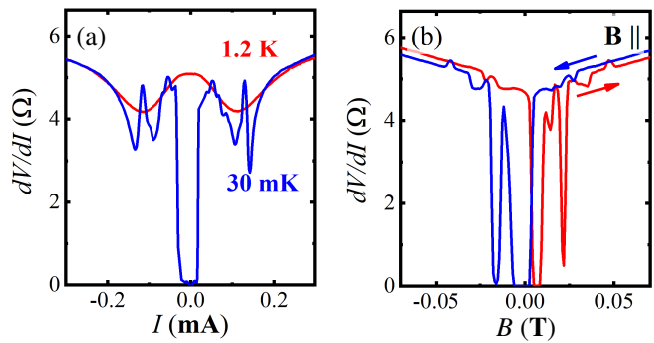}
\caption{(Color online) (a) Josephson current  for a double In-GeTe-In junction at 30~mK temperature (blue curve). Josephson current is suppressed above 1~K temperature, so the $dV/dI(I)$ differential resistance is of standard Andreev shape at 1.2~K temperature (red curve). The bulk indium superconducting gap 0.5~meV is seen as two resistance minima at $\approx\pm 0.1$~mA current bias (see the text). Due to the indium diffusion, the spin-polarized surface has no effect on the measured superconducting gap. (b) Suppression of the Josephson current by magnetic field for two opposite sweep directions at 30~mK temperature. The curves shows reentrant effect: when sweeping field from the superconducting (zero-resistance) state, Josephson current reappears at some finite magnetic field, i.e. around -0.02~T for the blue curve and around +0.02~T for the red curve.
}
\label{fig4}
\end{figure}

If the gap is suppressed due to the properties of GeTe surface~\cite{spin text,GeTesurfStates}, it can be easily affected by In diffusion. For this reason, the samples are heated for 3 minutes to 200$^\circ$ C after the Andreev reflection investigations. The In diffusion region is known to be around 100-200~nm in this case, so it is still one order of  magnitude smaller than the separation between the In leads. However, $dV/dI$ differential resistance is changed dramatically in Fig.~\ref{fig4}. 

At the lowest 30~mK temperature, the zero-bias anomaly drops to the zero resistance value, see Fig.~\ref{fig4} (a), so Josephson current connects the voltage probes. The zero-resistance  region is slightly asymmetric, which reflects usual Josephson current hysteresis with the sweep direction.  

Josephson current is suppressed above 1~K temperature in Fig.~\ref{fig4} (a), so the $dV/dI(I)$ curve is of standard Andreev shape~\cite{tinkham} at 1.2~K. There are two resistance minima at $\approx\pm 0.1$~mA current bias. The normal resistance value is $\approx 5\Omega$ in Fig.~\ref{fig4} (a), so the superconducting gap can be estimated as 
0.1~mA$\times$~5~$\Omega \approx 0.5$~meV, which well corresponds to the  bulk indium gap~\cite{indium}. This estimation confirms indium diffusion, so the spin-polarized surface has no effect on the measured superconducting gap. 

Fig.~\ref{fig4} (b) shows suppression of the Josephson current by magnetic field for two opposite sweep directions. In contrast to the Andreev curves in  Fig.~\ref{fig2} (b), prominent hysteresis is seen in Fig.~\ref{fig4} (b) with the field sweep direction. In addition, the curves demonstrate reentrant behavior of the Josephson current: when sweeping from the superconducting (zero-resistance) state, it is suppressed at some magnetic field value, while  reappears at higher fields, e.g. around 0.02~T for the blue curve in Fig.~\ref{fig4} (b).

\begin{figure}
\includegraphics[width=\columnwidth]{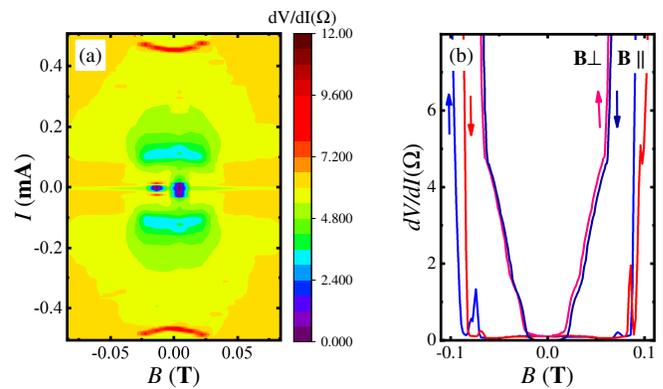}
\caption{(Color online) (a) Colormap, demonstrating the reentrant behavior of the Josephson current for the sample from Fig.~\ref{fig4}. In-plane magnetic field is changed from the positive values to the negative ones (as for the blue curve in Fig.~\ref{fig4} (b)), at every field $dV/dI(I)$ differential resistance is traced. The zero-resistance state forms two distinct regions, which are separated by the finite resistance values. (b) Reentrant behavior of the Josephson current for another sample. When sweeping from the zero-resistance state,  it  is suppressed and further reappears around $\pm$0.1~T.  For the normal magnetic field orientation, the critical current is much smaller, the $dV/dI(B)$ curves are nearly symmetric, no reentrant behavior is observed. All the data are  at 30~mK temperature. 
}
\label{fig5}
\end{figure}

The reentrant effect is also shown by the colormap in Fig.~\ref{fig5} (a). The in-plane magnetic field is changed from the positive values to the negative ones, at every field $dV/dI(I)$ differential resistance is traced. The zero-resistance state forms two distinct areas, which are separated by the finite resistance region, in  good correspondence to the scan in Fig.~\ref{fig4} (b). The asymmetry of the colormap is due to the magnetic field sweep direction, like in Fig.~\ref{fig4} (b).

Qualitatively similar reentrant behavior of the Josephson current can be observed for different samples, see  Fig.~\ref{fig5} (b) as an example of the In-GeTe-In device with maximum critical field value. For this sample, when sweeping from the zero-resistance state,  Josephson current is suppressed and further reappears around $\pm$0.1~T for the in-plane field. For the normal magnetic field orientation in Fig.~\ref{fig5} (b), the critical current is much smaller, the $dV/dI(B)$ curves are nearly symmetric, no reentrant behavior is observed.

\section{Discussion}

As a result, we have detected nonmonotonic effects of the applied external magnetic field in 
In-GeTe proximity devices, including reentrant superconductivity in In-GeTe-In Josephson junctions. In the latter case, supercurrent reappears at some finite magnetic field, see Figs.~\ref{fig4} and~\ref{fig5}. For the Andreev reflection, the superconducting gap is partially suppressed in zero magnetic field, while it is increased nearly to the bulk value for some finite field before its full suppression in Fig.~\ref{fig3}. 

Reentrant phase coherence is widely known in superconducting composites, see e.g. 
Ref.~\onlinecite{reentrant-phase} and references therein. In these highly disordered systems, it is usually attributed to electron localization/delocalization processes. This mechanism can not be applied to our homogeneous GeTe single crystal samples. There is also a possibility~\cite{hotelectrons} to observe nonmonotonic effect of magnetic field on the switching current of Josephson junctions with spatially separated current and voltage probes due to improved quasiparticle thermalization by a magnetic field. For our samples, the grounded contact is always used as one of the potential probes, which removes the problem of thermalization. Moreover, we do not observe any difference for various contact configurations. Further,  reentrant superconductivity with varying external magnetic field was predicted to 
take place in homogeneous samples of layered quasi-two-dimensional triplet 
superconductors~\cite{Lebed1998,Lebed1999,Mineev2000,Mineev2020,Lebed2020}. However, one of the main assumptions of that quasi-two dimensional model, that the superconductor coherence length is less or comparable with the interlayer distance, is violated for  indium and GeTe.

Since indium is a conventional s-wave superconductor, the observed effects should be mostly associated with specific properties of the proximized noncentrosymmeric (ferroelectric) topological semimetal $\alpha$-GeTe~\cite{ortix,triple-point}. 

First of all, the undoped GeTe is not a superconductor~\cite{indium-doping,Mn-doping}. For our single crystals, the powder X-ray diffraction analysis confirms pure single-phase GeTe, also, GeTe composition is verified by energy-dispersive X-ray spectroscopy. Even after the sample's heating in Sec.~\ref{double}, the In diffusion region is about 100-200~nm, so even the GeTe surface is not superconducting in 5~$\mu$m regions between the In leads.

There are several characteristic features of $\alpha$-GeTe, which seems to  be important in 
Nb-GeTe proximity devices.  First of all, topological surface states with nontrivial spin textures~\cite{triple-point,GeTesurfStates} can play an important role. In addition, there are nontrivial spin textures with nonzero spin winding numbers in the bulk of $\alpha$-GeTe, which are associated with the type-II Weyl fermions around the triple points of the electronic band structure~\cite{triple-point}. Also, the pronounced spin-orbit splitting~\cite{GeTesurfStates,triple-point} can influence on the proximity-induced odd-frequency triplet component of the superconductor order parameter and on the charge transport as a whole.

Thus, it seems to be quite natural to model In-GeTe heterostructures as SFN junctions with some definite magnetization and  strong spin-orbit coupling. Inhomogeneous spin directions, that are possibly incorporated in the textures, could complicate the model, reduce its anisotropic properties and contribute to the field-induced nonmonotonic behavior~\cite{BBK2002,Sperstadetal2008,Mengetal2013}.

As for the Andreev reflection, the bulk indium gap is partially suppressed  to 0.25~meV in Fig.~\ref{fig2} (a) due to the spin polarization of the GeTe surface states~\cite{GeTesurfStates}. This suppression disappears after indium diffusion, see  the red curve in Fig.~\ref{fig4} (a), which confirms the important role of the surface states and the inverse proximity effect as the origin of the zero-field gap suppression in Figs.~\ref{fig2} and~\ref{fig3}.  The external magnetic field could modify  the spin configuration, which firstly increases the 
superconducting gap to nearly the bulk 0.5~meV value, and suppresses it to zero  in higher fields, see Fig.~\ref{fig3}.  For the Josephson effect, supercurrent flows along the GeTe surface between the In leads in the planar experimental geometry, see  Fig.~\ref{fig1}. Together with the gap value, the initial  spin polarization of surface states partially suppresses the critical current, which is recovered due to the modified spin-split surface states in finite field before the full Josephson effect suppression in Figs.~\ref{fig4} and~\ref{fig5}. 

Thus, we attribute the order parameter depletion (or healing) to the consequence of the inverse proximity effect in superconductor-ferromagnet structures~\cite{Bergeretetal2004,Bergeretetal2005,Kharitonovetal2006,Linderetal2009,Greinetal2013,MironovMelnikovBuzdin2018,Volkovetal2019,Suzukietal2021,Mironovetal2021,Flokstraetal2021,Melnikovetal2022}. 
In topological materials, the inverse proximity effect is defined by the relation between the proximity-induced singlet and triplet correlations, which  can  generally be tuned by varying external magnetic field~\cite{Bursetetal2015}. In superconductor-Weyl semimetal junctions, depletion of the order parameter can be very sensitive to relative orientations between the applied magnetic field (weaker than the superconductor critical field) and the surface magnetization~\cite{Mohantaetal2020}. The relative orientations can also control the existence of zero-energy peak in the local density of states in hybrid SF systems in the presence of the exchange field together with the spin-orbit coupling in the ferromagnet~\cite{Jacobsenetal2015}. Moreover, a pronounced nonmonotonic behavior of the critical temperature as a function of the exchange field direction has been predicted in such SF bilayers~\cite{Jacobsenetal2015}. 
In our experiment, the total Zeeman field, formed by noncollinear external magnetic and exchange fields, should experience some effective rotation with changing the applied field value.

The proposed qualitative description is to some extent similar to the field-induced reentrant effects, which have  been identified in uniform materials  with combined magnetization and
strong spin-orbit coupling,  such as the uranium-based superconductors~\cite{Levyetal2005,Miyakeetal2008,Mineev2017,Machida2021,Kinjoetal2023,Rosueletal2023}.

However, more detailed theoretical studies are required for an unambiguous interpretation of the experimental results. For example, nonmonotonic field dependence of the critical temperature occurs after the transition into the Fulde-Ferrell-Larkin-Ovchinnikov (FFLO) state~\cite{Mironovetal2012,Mironovetal2018,MarychevVodolazov2018,MarychevPlastovetsVodolazov2020} 
(for reviews, see Refs.~\onlinecite{Mironovetal2021} and~\onlinecite{Melnikovetal2022}) in  hybrid layered  SFN heterostructures. The FFLO physics is based on Cooper pairing with finite momentum against a background of the Zeeman splitting, which usually leads to damped oscillations of the Cooper pair wave function across the ferromagnetic 
layer~\cite{BBP1982,Jiangetal1995,Fominovetal2002,Zdravkovetal2006,Sidorenko2009,buzdin2005}. 
When SFN threelayers are in the FFLO state and the oscillations are along the layers, the penetration depth and the critical temperature can manifest a nonmonotonic magnetic field dependence, that can result in a reentrant behavior with 
varying in-plane magnetic field~\cite{MarychevVodolazov2018}. For the SFN strip, the nonmonotonic 
behavior can  take place, when the magnetic field is perpendicular to the interfaces~\cite{MarychevPlastovetsVodolazov2020}. 
The order parameter and the critical temperature in SF bilayer can be enhanced with increasing the in-plane field in the FFLO state~\cite{Bobkovs2014,MarychevVodolazov2018}. In the SFNFS pentalayers, the transitions from the $\pi$-state to the $0$-state  can also occur~\cite{MarychevVodolazov2018}.  Although the FFLO scenario can generally result in the reentrant behavior and looks potentially promising, the role of the strong spin-orbit coupling has not been analyzed in these effects yet, which is crucial for GeTe topological semimetal.

\section{Conclusion}

As a conclusion, we  experimentally investigate charge transport in In-GeTe and In-GeTe-In 
proximity devices and observe nonmonotonic effects of the applied external magnetic field, 
including reentrant superconductivity in In-GeTe-In Josephson junctions.  We discuss possible 
reasons for the results obtained, taking into account  spin polarization of Fermi arc surface 
states in topological semimetal $\alpha$-GeTe with a strong spin-orbit coupling.  In particular, 
the zero-field surface state spin polarization in the the proximity-influenced system partially 
suppresses the superconductivity over the scale of the superconducting coherence length, while it 
is recovered due to the nonmonotonic inverse proximity effects and/or modified spin-split surface 
state configuration in finite fields before the full suppression above the critical field. As an 
alternative possible scenario, the transition into the Fulde-Ferrell-Larkin-Ovchinnikov state is 
discussed, which can generally result in the reentrant behavior. However, the role of  strong 
spin-orbit coupling has not been analyzed theoretically for heterostructures in the 
Fulde-Ferrell-Larkin-Ovchinnikov state, which is crucial for junctions involving GeTe topological 
semimetal.

\acknowledgments
We wish to thank S.S.~Khasanov for X-ray sample characterization. We gratefully acknowledge financial support  by the  Russian Science Foundation, project RSF-22-22-00229, https://rscf.ru/project/22-22-00229/.

\end{document}